\documentclass[sigconf]{acmart}

\AtBeginDocument{%
  }

\copyrightyear{2025}
\acmYear{2025}
\setcopyright{rightsretained}
\acmConference[RecSys '25]{Proceedings of the Nineteenth ACM Conference on Recommender Systems}{September 22--26, 2025}{Prague, Czech Republic}
\acmBooktitle{Proceedings of the Nineteenth ACM Conference on Recommender Systems (RecSys '25), September 22--26, 2025, Prague, Czech Republic}\acmDOI{10.1145/3705328.3759310}
\acmISBN{979-8-4007-1364-4/2025/09}

\definecolor{mycyan}{rgb}{0.0, 1.0, 1.0}     
\definecolor{myteal}{rgb}{0.0, 0.5, 0.5}

\usepackage{amsmath}
\usepackage{amsfonts}
\usepackage{mathtools}
\usepackage{graphicx}
\usepackage{booktabs} 
\usepackage[inline]{enumitem}
\usepackage{balance}
\usepackage[skip=0pt]{caption}
\usepackage{multirow}
\usepackage{siunitx}
\usepackage{acronym}
\usepackage{algorithm}
\usepackage{subcaption}

\usepackage{csvsimple}

\usepackage{pgfplots}
\usepackage{tikz}
\usetikzlibrary{calc}

\usepackage{xspace}
\usepackage{booktabs,multirow,siunitx}
\definecolor{rowgray}{gray}{0.93}
\newcommand{\algname}[1] {{\fontfamily{cmtt}\selectfont {#1}}}

\begin{document}




\title{Opening the Black Box: Interpretable Remedies for Popularity Bias in Recommender Systems}

\author{Parviz Ahmadov}
\orcid{0009-0001-5334-1920}
\affiliation{%
  \institution{Delft University of Technology}
  \city{Delft}
  \country{The Netherlands}
}
\email{p.ahmadov@student.tudelft.nl}

\author{Masoud Mansoury}
\orcid{0000-0002-9938-0212}
\affiliation{%
  \institution{Delft University of Technology}
  \city{Delft}
  \country{The Netherlands}
}
\email{m.mansoury@tudelft.nl}

\begin{abstract}

Popularity bias is a well-known challenge in recommender systems, where a small number of popular items receive disproportionate attention, while the majority of less popular items are largely overlooked. This imbalance often results in reduced recommendation quality and unfair exposure of items. Although existing mitigation techniques address this bias to some extent, they typically lack transparency in how they operate. In this paper, we propose a post-hoc method using a Sparse Autoencoder (SAE) to interpret and mitigate popularity bias in deep recommendation models. The SAE is trained to replicate a pre-trained model’s behavior while enabling neuron-level interpretability. By introducing synthetic users with clear preferences for either popular or unpopular items, we identify neurons encoding popularity signals based on their activation patterns. We then adjust the activations of the most biased neurons to steer recommendations toward fairer exposure. Experiments on two public datasets using a sequential recommendation model show that our method significantly improves fairness with minimal impact on accuracy. Moreover, it offers interpretability and fine-grained control over the fairness–accuracy trade-off.
 
\end{abstract}



\maketitle

\section{Introduction}

Popularity bias is a persistent challenge in recommender systems, where highly popular items are favored over less popular, long-tail items that may be more relevant to niche user interests~\cite{canamares2018should, zhang2023empowering, liu2020long}. This bias arises naturally from collaborative filtering and other learning-based approaches that reflect patterns in historical interaction data, including skewness toward popular content~\cite{schnabel2016recommendations,huang2024going,mansoury2020feedback}. While such recommendations may serve mainstream users well, they limit opportunities for discovery, marginalize niche or newer items, and reduce exposure fairness. This can ultimately disadvantage both content creators and users with specialized tastes~\cite{zhu2022fighting,greenwood2024user,mansoury2022understanding,mansoury2020feedback}.

Most existing methods to mitigate popularity bias rely on reweighting item scores or modifying model architectures~\cite{zhu2021popularity,chen2021autodebias,mansoury2024mitigating}. However, these techniques often function as black boxes, offering limited insight into why certain items are recommended or which internal components drive biased behavior. This lack of interpretability poses a challenge for diagnosing and correcting systemic biases.

To address this, we turn to Sparse Autoencoders (SAEs)—neural models designed to activate only a small subset of neurons per input. This sparsity enables clearer attribution of individual neurons to specific features or concepts. Recent work in interpretability has shown that SAEs can uncover high-level, monosemantic neurons in large language models~\cite{bricken2023monosemanticity, gao2024scaling, o2024disentangling, lan2024sparse}. Crucially, sparsity makes these neurons easier to isolate and manipulate without unintended side effects, a property already leveraged to mitigate bias in language models~\cite{durmus2024steering, hegde2024effectiveness, harle2024scar}.

Inspired by these findings, we propose \algname{PopSteer}, a novel post-hoc method that interprets and mitigates popularity bias at the neuron level in deep recommendation models. We begin by training an SAE to replicate the output of a pretrained recommender, attaching it to the final layer to capture the model’s decision-making process. Then, we generate synthetic user profiles that reflect strong preferences for either popular or unpopular items. By analyzing how individual SAE neurons respond to these inputs, we identify those most responsible for encoding popularity signals.

Once identified, \algname{PopSteer} adjusts neuron activations to reduce the influence of popularity-biased neurons and amplify the contribution of neurons aligned with long-tail content. This neuron steering approach enables fine-grained control over recommendation behavior while preserving the model's overall structure and performance. Our experiments on two public datasets using the SASRec model demonstrate that \algname{PopSteer} effectively improves item exposure fairness with minimal loss in recommendation accuracy. Compared to several existing bias mitigation techniques, our method offers both superior performance and interpretability.

\section{Interpretation with Sparse Autoencoder}\label{sec:interpret_SAE}

Formally, let $x \in \mathbb{R}^d$ denote an embedding from the ML model. The SAE, consisting of an input dimension of size $d$, a hidden dimension of size $N$, and an output dimension of size $d$, reconstructs $x$ as:
\begin{equation*} 
\hat{x} = W_{\text{dec}} a + b_{\text{pre}} 
\end{equation*}
\noindent where $W_{\text{dec}} \in \mathbb{R}^{d \times N}$ and $b_{\text{pre}} \in \mathbb{R}^d$ are learnable parameters. The hidden representation $a \in \mathbb{R}^N$ is computed as:
\begin{equation*} 
a = \text{ReLU}(z) 
\end{equation*}
\begin{equation*} 
z = W^T_{\text{enc}} (x - b_{\text{pre}}) 
\end{equation*} 
\noindent where $W_{\text{enc}} \in \mathbb{R}^{d \times N}$ is the learnable encoder weight matrix. While ReLu introduces basic sparsity, stricter control is applied by retaining only the top-$K$ highest activations in $a$ and setting all other activations to zero, as proposed by Gao et al.~\cite{gao2024scaling}. With $K<d$, the top-$K$ mask activates just a limited subset of neurons. Thanks to the large hidden dimension $N$, the model can pick from many candidates, and the resulting sparsity encourages each fired unit to \emph{specialize} in a meaningful feature of the input. The SAE is trained to minimize the following objective: 
\begin{equation} 
\min_{W_{\text{enc}}, W_{\text{dec}}, b_{\text{pre}}} \| x - \hat{x} \|_2^2 + \gamma * \mathcal{L}_{aux} 
\end{equation}
\noindent where $\mathcal{L}_{aux}$ is a loss term preventing dead neurons\footnote{We address the dead neuron issue during SAE training using the auxiliary loss, but omit the details here due to space constraints (see~\cite{gao2024scaling} for more).}.
This objective ensures that the sparse hidden representation maintains fidelity to the original embedding while revealing the underlying decision structure. The learned neurons can then be interpreted to explain which concepts in the input data influence the ML model's output.

\section{Interpreting Popularity Bias}

Since each SAE neuron tends to specialize in a distinct concept from the input data, analyzing their activation behavior reveals how such concepts influence predictions. To explore this, we generate synthetic user profiles that highlight popularity bias and examine how specific neurons respond. This enables the identification of neurons that either reinforce or counteract popularity signals. 

Each synthetic profile is passed through the pretrained recommendation model to obtain the user embedding (i.e., $x$ in section~\ref{sec:interpret_SAE}), which is then fed into the pretrained SAE for neuron-level analysis. In the following, we describe our synthetic data generation process and how we quantify each neuron's contribution to popularity bias.

\subsection{Generating synthetic datasets}

We create two types of synthetic user profiles: one biased toward popular items and the other toward unpopular items. These profiles simulate user interactions that reflect extreme cases of popularity preference, enabling clearer observation of neuron activation.

Let $\mathcal{U} = \{u_1, ..., u_n\}$ be the set of users, $\mathcal{I} = \{i_1, ..., i_m\}$ be the set of items, and $R \in \mathbb{R}^{n \times m}$ represent the user-item interaction matrix. Following~\cite{abdollahpouri2021user,zhang2023empowering}, we define $\mathcal{I}^{pop}$ (popular or \textit{head} items) as the most frequently interacted items and $\mathcal{I}^{unpop}$ (unpopular or \textit{tail} items) as the least interacted items, both compromising roughly 20\% of total interactions. Using these popular and unpopular item sets, we construct two synthetic datasets: 1) $R^{pop}$ with user profiles containing only popular items, and 2) $R^{unpop}$ with user profiles containing only unpopular items.

To generate $R^{pop}$, for each user $u$ in $R$, we follow this process: 1) extract the items interacted by $u$ in $R$, 2) replace each item $i$ with a randomly selected item from $\mathcal{I}^{pop}$, and 3) add this modified profile to $R^{pop}$. We apply the same procedure for generating $R^{unpop}$, using items from $\mathcal{I}^{unpop}$ in step 2. 
This approach preserves the number of interactions per user, isolating the popularity signal without altering profile length or interaction density. Importantly, these synthetic datasets are not used to train the SAE. Instead, we feed them into a pretrained SAE (trained on $R$), to evaluate its neurons' activation, ensuring that its learned representations remain grounded in real user behavior.

\subsection{Detecting Neurons Encoding Popularity Bias}

To quantify how much each SAE neuron contributes to popularity bias, we compare neuron activations when processing $R^{pop}$ and $R^{unpop}$. Each synthetic dataset is fed into the pretrained SAE, and the activation levels of the hidden layer neurons are recorded.

Prior work suggests that as neural networks grow in size, neuron activation distributions tend to approximate a Gaussian distribution~\cite{lee2017deep, wolinski2022gaussian, haider2025neurons}. 
We leverage this property to use Cohen’s $d$~\cite{cohen2013statistical} to measure the effect size of activation differences for each neuron $j$:
\begin{equation} 
d_j = \frac{\mu_{j,pop} - \mu_{j,unpop}}{\sqrt{\frac{\sigma^2_{j,pop} + \sigma^2_{j,unpop}}{2}}} 
\end{equation}
\noindent where $\mu_{j,pop}$ and $\mu_{j,unpop}$ are the mean activations of neuron $j$ under $R^{pop}$ and $R^{unpop}$, respectively, and $\sigma_{j,pop}$ and $\sigma_{j,unpop}$ are the corresponding standard deviations.

Cohen’s $d$ provides a normalized measure of the difference between two distributions. A high absolute value of $d_j$ indicates that neuron $j$ is strongly responsive to popularity-related patterns. Positive values suggest alignment with popular content, while negative values imply a focus on unpopular or niche items.

\begin{figure*}[htbp]
\centering
    \newcommand{\Width}{0.95\columnwidth}
    \newcommand{\Height}{0.65\columnwidth}

    \begin{subfigure}[b]{0.49\textwidth}
        \centering
        \begin{tikzpicture}
        \begin{axis}[
            xlabel=nDCG,
            ylabel=Long-tail coverage,
            width=\Width,
            height=\Height,
            legend pos=north west,
            legend cell align={left},
            legend style={fill opacity=0.75, draw opacity=1, text opacity=1},
            x label style={yshift=0.5em},
            y label style={yshift=-0.5em},
        ]
        \addplot[
            scatter, only marks, scatter src=explicit symbolic,
            scatter/classes={
                SASRec={mark=x, color={rgb, 255:red, 0; green, 128; blue, 128},mark size=3,line width=1.1pt},
                Random={mark=star, color={rgb, 255:red, 255; green, 182; blue, 193},mark size=2.5,line width=1.1pt},
                PCT={mark=diamond*, fill=purple!40!white,mark size=2.5},
                PMMF={mark=triangle*, fill=green!40!white,mark size=2.5},
                IPR={mark=square*, fill=yellow!40!white},
                FA*IR={mark=*, fill=orange!40!white},
                PopSteer={mark=square*, fill=blue!40!white}
            }
        ]
        table [x=ndcg, y=Deep long tail coverage, col sep=comma, meta=approach] {baselines_ml.csv};
        \legend{SASRec,Random,PCT,P-MMF,IPR,FA*IR,PopSteer}
        \end{axis}
        \end{tikzpicture}
    \end{subfigure}
    \begin{subfigure}[b]{0.49\textwidth}
        \centering
        \begin{tikzpicture}
        \begin{axis}[
            xlabel=nDCG,
            ylabel=Gini Index,
            width=\Width,
            height=\Height,
            legend pos=north west,
            legend cell align={left},
            legend style={fill opacity=0.75, draw opacity=1, text opacity=1},
            x label style={yshift=0.5em},
            y label style={yshift=-0.5em},
        ]
        \addplot[
            scatter, only marks, scatter src=explicit symbolic,
            scatter/classes={
                PopSteer={mark=square*, fill=blue!40!white},
                FA*IR={mark=*, fill=orange!40!white},
                IPR={mark=square*, fill=yellow!40!white},
                PCT={mark=diamond*, fill=purple!40!white,mark size=2.5},
                PMMF={mark=triangle*, fill=green!40!white,mark size=2.5},
                Random={mark=star, color={rgb, 255:red, 255; green, 182; blue, 193},mark size=2.5,line width=1.1pt},
                SASRec={mark=x, color={rgb, 255:red, 0; green, 128; blue, 128},mark size=3,line width=1.1pt}
            }
        ]
        table [x=ndcg, y=gini, col sep=comma, meta=approach] {baselines_ml.csv};
        \end{axis}
        \end{tikzpicture}
    \end{subfigure}

\caption{Performance comparison of \algname{PopSteer} method with the baselines in terms of nDCG and fairness metrics on ML-1M.}
\label{fig:baseline_ml}
\end{figure*}

\begin{figure*}[htbp]
\centering
    \newcommand{\Width}{0.95\columnwidth}
    \newcommand{\Height}{0.65\columnwidth}

    \begin{subfigure}[b]{0.49\textwidth}
        \centering
        \begin{tikzpicture}
        \begin{axis}[
            xlabel=nDCG,
            ylabel=Long-tail coverage,
            width=\Width,
            height=\Height,
            legend pos=north west,
            legend cell align={left},
            legend style={fill opacity=0.75, draw opacity=1, text opacity=1},
            x label style={yshift=0.5em},
            y label style={yshift=-0.5em},
        ]
        \addplot[
            scatter, only marks, scatter src=explicit symbolic,
            scatter/classes={
                SASRec={mark=x, color={rgb, 255:red, 0; green, 128; blue, 128},mark size=3,line width=1.1pt},
                Random={mark=star, color={rgb, 255:red, 255; green, 182; blue, 193},mark size=2.5,line width=1.1pt},
                PCT={mark=diamond*, fill=purple!40!white,mark size=2.5},
                PMMF={mark=triangle*, fill=green!40!white,mark size=2.5},
                IPR={mark=square*, fill=yellow!40!white},
                FA*IR={mark=*, fill=orange!40!white},
                PopSteer={mark=square*, fill=blue!40!white}
            }
        ]
        table [x=ndcg, y=Deep long tail coverage, col sep=comma, meta=approach] {baselines_lastfm.csv};
        \legend{SASRec,Random,PCT,P-MMF,IPR,FA*IR,PopSteer}
        \end{axis}
        \end{tikzpicture}
    \end{subfigure}
    \begin{subfigure}[b]{0.49\textwidth}
        \centering
        \begin{tikzpicture}
        \begin{axis}[
            xlabel=nDCG,
            ylabel=Gini Index,
            width=\Width,
            height=\Height,
            legend pos=north west,
            legend cell align={left},
            legend style={fill opacity=0.75, draw opacity=1, text opacity=1},
            x label style={yshift=0.5em},
            y label style={yshift=-0.5em},
        ]
        \addplot[
            scatter, only marks, scatter src=explicit symbolic,
            scatter/classes={
                PopSteer={mark=square*, fill=blue!40!white},
                FA*IR={mark=*, fill=orange!40!white},
                IPR={mark=square*, fill=yellow!40!white},
                PCT={mark=diamond*, fill=purple!40!white,mark size=2.5},
                PMMF={mark=triangle*, fill=green!40!white,mark size=2.5},
                Random={mark=star, color={rgb, 255:red, 255; green, 182; blue, 193},mark size=2.5,line width=1.1pt},
                SASRec={mark=x, color={rgb, 255:red, 0; green, 128; blue, 128},mark size=3,line width=1.1pt}
            }
        ]
        table [x=ndcg, y=gini, col sep=comma, meta=approach] {baselines_lastfm.csv};
        \end{axis}
        \end{tikzpicture}
    \end{subfigure}

\caption{Performance comparison of \algname{PopSteer} method with the baselines in terms of nDCG and fairness metrics on Last.fm.}
\label{fig:baseline_lastfm}
\end{figure*}

\section{Popularity Bias Mitigation Approach} \label{popularity_mitigation}

To counteract popularity bias in recommendations, we use the computed Cohen's d values to identify neurons most responsible for encoding popularity signals. Our method, \algname{PopSteer}, applies a process called \emph{neuron steering}, which systematically adjusts these neuron activations to reduce bias and promote fairer item exposure.

Neuron steering modifies the hidden activations within the SAE to either amplify or suppress the influence of popularity-related neurons. The adjustment is guided by each neuron's $d_j$ score, which reflects its alignment with popular or unpopular items. We first assign a weight $w_j$ to each neuron $j$ based on the normalized absolute value of its $d_j$ score. This weight determines how much each neuron's activation will be adjusted:
\begin{equation}\label{eq_weight_alpha}
w_j = \alpha \cdot \frac{|d_j| - \min(|d|)}{ \max(|d|) - \min(|d|) } 
\end{equation}
\noindent where $\alpha$ is a tunable hyperparameter that controls the overall strength of the steering and $d$ is the Cohen's d values of all neurons. Neurons with stronger associations to popularity bias (higher $|d_j|$) receive larger weights. Next, we modify the original activation $a_j$ of neuron $j$ to obtain a new activation $a_j^\prime$: 

\begin{equation} 
a_j^\prime = 
\begin{cases} 
a_j + w_j \cdot \sigma_j, & \text{if } d_j < 0 \\ 
a_j - w_j \cdot \sigma_j, & \text{if } d_j > 0 
\end{cases} 
\end{equation}

This adjustment boosts neurons promoting unpopular items ($d_j < 0$) and suppresses those favoring popular items ($d_j >0$). By realigning neuron activations, \algname{PopSteer} mitigates the overrepresentation of popular items, leading to more balanced exposure across items. The updated SAE then returns the modified user embedding, which is used alongside item embeddings from the base recommendation model to generate the final recommendation list. 

\section{Experiments}

This section outlines our experimental setup, including datasets, evaluation metrics, baselines, and experimental results.

\subsection{Dataset}
We conduct experiments on two public datasets: MovieLens 1M (ML-1M)~\cite{harper2015movielens} and Last.fm~\cite{schedl2016lfm}. Following standard preprocessing~\cite{he2017translation, kang2018self}, we create 5-core sample on both datasets, where each user has at least 5 interactions and each item is interacted with by at least 5 users. Both datasets include timestamp informtion, enabling their use in sequential recommendation task. Table~\ref{table_datastats} summarizes the statistical properties of these datasets.

\begin{table}[t!]
  \centering
  \caption{Statistics of the datasets.}
  \begin{tabular}{lrrrr}
    \toprule
    Dataset & \#Users & \#Items & \#Interactions & Density \\
    \midrule
    ML-1M    & 6,040    & 3,417    & 999,611        & 0.048 \\
    Last.fm  & 1,363    & 2,976    & 64,286         & 0.016 \\
    \bottomrule
    \label{table_datastats}
  \end{tabular}
\end{table}

\subsection{Experimental setup}

We evaluate recommendation performance along two dimensions: ranking performance and fairness. Ranking quality is measured using nDCG@10. To evaluate fairness, we report long-tail coverage~\cite{mansoury2022understanding}—the proportion of recommendations that come from the unpopular (tail) item set ($\mathcal{I}^{unpop}$)—and the Gini Index~\cite{antikacioglu2017post}, which captures how uniformly items are exposed across users. A lower Gini Index indicates fairer exposure.

We compare our \algname{PopSteer} method with the following baselines:
\begin{itemize}
    \item \textbf{Inverse Popularity Ranking (IPR) \cite{zhang2010niche}:} A reweighting approach that adjusts relevance score for user-item pairs based on inverse popularity: $\tilde{s}_{u,i}= s_{u,i}/\!\left(1+\alpha\,\rho_i\right)$, where $\rho_i=\mathrm{pop}(i)\big/\max_{j\in\mathcal{I}}\mathrm{pop}(j)$ and $\alpha$ is a hyperparameter controlling the degree of mitigating bias. We tune $\alpha \in \{0.1\allowbreak,0.3\allowbreak,0.5\allowbreak,0.7\allowbreak,1\}$.
    \item \textbf{FA*IR \cite{zehlike2017fa}:} A post-processing approach that enforces fairness by ensuring a minimum proportion of unpopular items appear within each prefix of the top-k list. In our experiment, we set the proportion of unpopular items to $p \in \{0.3\allowbreak,0.5\allowbreak,0.7\allowbreak,0.9\allowbreak,0.99\}$ and the significance level to $\alpha \in \{0.01\allowbreak,0.05\allowbreak,0.1\}$. We set the size of long recommendation lists to 500.
    \item \textbf{PCT \cite{wang2023two}:} A two-sided method that balances user-level and system-level exposure. A solver computes exposure targets using linear programming, followed by a reranker that modifies each user’s recommendation list using a modified MMR strategy. Hyperparameters are tuned similar to FA*IR. 
    \item \textbf{P-MMF \cite{xu2023p}:} A resource allocation algorithm based on dual-space optimization. It dynamically and proportionally adjusts exposure across popular, unpopular, and mid items according to group sizes, updating after each user interaction. We tune the learning rate $\eta \in \{0.0001\allowbreak,0.001\allowbreak,0.01\}$ and the fairness constraint parameter $\lambda \in \{0.001\allowbreak, 0.01\allowbreak, 0.1\allowbreak, 1\allowbreak, 10\}$.
    \item \textbf{Random:} A simple baseline that randomly selects $k$ items from an initial longer recommendation list. We test the method using list sizes $\{15\allowbreak,30\allowbreak,50\allowbreak,75\allowbreak,100\}$.
\end{itemize}

\begin{figure}[t]
    \centering
    \includegraphics[width=\linewidth]{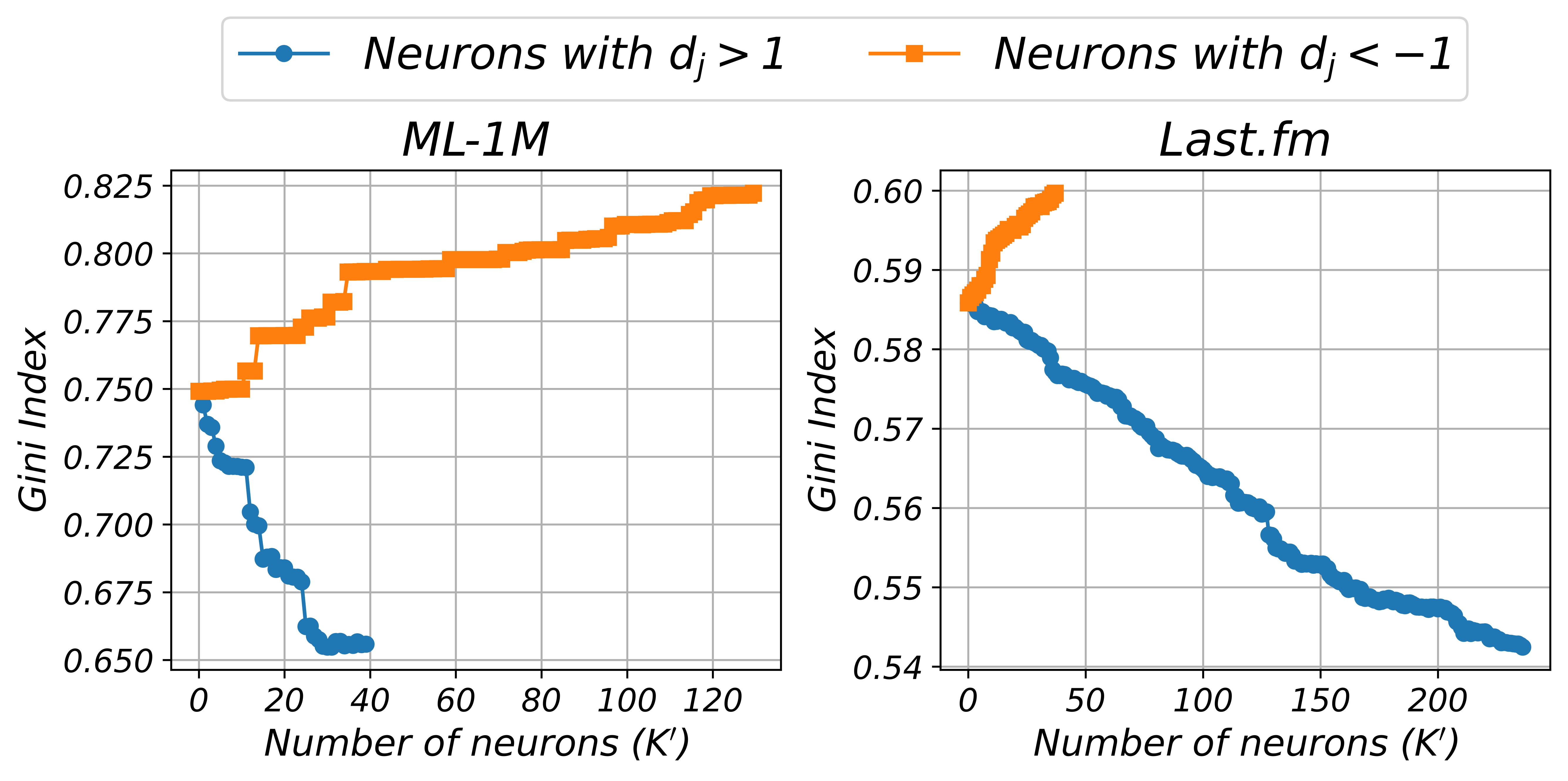}
    \caption{Interpretability analysis of \algname{PopSteer}: effect of deactivating $K^\prime$ identified neurons linked to popularity bias.}\label{fig:interpretation}
\end{figure}

We use SASRec~\cite{kang2018self} as the backbone recommender, which leverages self-attention to capture temporal dynamics in user behavior. For splitting the data, user histories are sorted chronologically, with the most recent interaction used for testing, the second-most for validation, and the remainder for training.

SASRec is trained with a learning rate of 0.001 and early stopping (patience = 10 epochs) based on validation nDCG@10. For SAE, we use total reconstruction loss as the objective~\cite{gao2024scaling}. The SAE includes two key hyperparameters: the scale factor $s \in \{8, 32, 64\}$, which determines the size of the hidden layer relative to the input, and the sparsity level $K \in \{16, 32, 48\}$, which specifies the number of top activations retained in the hidden layer. Both SASRec and SAE are trained with the Adam optimizer and a batch size of 4096.

Our proposed \algname{PopSteer} method involves two tunable hyperparameters. The first is $\alpha$, which controls the intensity of neuron steering (see Eq.~\ref{eq_weight_alpha}), which we set to  $\alpha \in \{1.0, 1.5, 2.0, 2.5, 3.0, 3.5, 4.0\}$. The second is $\mathcal{N}$, which specifies how many neurons with the highest absolute Cohen's d value are selected for activation adjustment, as detailed in Section~\ref{popularity_mitigation}. We tune $\mathcal{N}  \in \{1024, 2048, 3072, 4096\}$ out of total neurons $N=4096$. Our implementation and code is available at \href{Our proposed}{https://github.com/parepic/PopSteer}.

\subsection{Experimental results} 

Figures~\ref{fig:baseline_ml} and~\ref{fig:baseline_lastfm} present our main results, comparing \algname{PopSteer} with SASRec and several baselines across both nDCG and fairness metrics on the ML-1M and Last.fm.

On ML-1M (Figure~\ref{fig:baseline_ml}), \algname{PopSteer} consistently outperforms all baselines in enhancing fairness while maintaining strong accuracy. P-MMF achieves moderate fairness improvements but at the cost of notable accuracy loss and limited flexibility. While PCT, IPR, and FA*IR show performance comparable to \algname{PopSteer} near nDCG$\sim$0.12, two key patterns highlight \algname{PopSteer}'s superiority.

First, \algname{PopSteer} achieves a more favorable trade-off between fairness and accuracy, delivering significantly better fairness gains for only slight reductions in nDCG. Second, other baselines often continue to lose accuracy without corresponding fairness gains, particularly below nDCG$\sim$0.11. In contrast, \algname{PopSteer} maintains stability—avoiding unnecessary accuracy degradation when fairness stabilizes. This behavior suggests the reliability of \algname{PopSteer}.

A similar trend appears on Last.fm (Figure~\ref{fig:baseline_lastfm}). Although FA*IR achieves marginally higher long-tail coverage in some settings, \algname{PopSteer} consistently yields better Gini Index scores, indicating a more balanced distribution of item exposure. Overall, \algname{PopSteer} demonstrates greater reliability, consistency, and stability in balancing fairness and recommendation quality.

\begin{table}[t]
  \caption{Ablation study: \algname{PopSteer} (PS) vs.\ Random Noisification (ML-1M: $\alpha = 1.5$, $\xi = 0.35$; Last.fm: $\alpha = 4.0$, $\xi = 1.0$).}
  \label{tab:pop_vs_noise_core}
  \centering
  \small
  \setlength{\tabcolsep}{4pt}
  \begin{tabular}{l r cc cc cc}
    \toprule
        &         & \multicolumn{2}{c}{NDCG@10~$\uparrow$} &
                      \multicolumn{2}{c}{LT Cov.@10~$\uparrow$} &
                      \multicolumn{2}{c}{Gini@10~$\downarrow$} \\
    \cmidrule(lr){3-4}\cmidrule(lr){5-6}\cmidrule(lr){7-8}
    Dataset & $\mathcal{N}$ &
    PS & Noise &
    PS & Noise &
    PS & Noise \\
    \midrule
    \multirow{5}{*}{ML-1M}
      & 0    & 0.1166 & 0.1166 & 0.4917 & 0.4917 & 0.7491 & 0.7491 \\
      & 1024 & \textbf{0.1169} & 0.1123 & 0.5065 & \textbf{0.5333} & 0.7470 & \textbf{0.7387} \\
      & 2048 & \textbf{0.1167} & 0.1129 & \textbf{0.5346} & 0.5440 & \textbf{0.7315} & 0.7376 \\
      & 3072 & \textbf{0.1135} & 0.1103 & \textbf{0.5704} & 0.5463 & \textbf{0.7016} & 0.7367 \\
      & 4096 & 0.1106 & \textbf{0.1110} & \textbf{0.6241} & 0.5485 & \textbf{0.6677} & 0.7394 \\
    \midrule
    \multirow{5}{*}{Last.fm}
      & 0    & {0.6247} & 0.6247 & 0.8722 & 0.8722 & 0.5859 & 0.5859 \\
      & 1024 & \textbf{0.6186} & 0.5867 & \textbf{0.9015} & 0.8575 & \textbf{0.5193} & 0.5819 \\
      & 2048 & \textbf{0.6126} & 0.5932 & \textbf{0.9167} & 0.8598 & \textbf{0.4876} & 0.5885 \\
      & 3072 & \textbf{0.6044} & 0.5977 & \textbf{0.9161} & 0.8677 & \textbf{0.4840} & 0.5824 \\
      & 4096 & \textbf{0.6036} & 0.6015 & \textbf{0.9150} & 0.8628 & \textbf{0.4826} & 0.5878 \\
    \bottomrule
  \end{tabular}
\end{table}

\subsubsection{Interpretability Analysis}

To assess the interpretability of \algname{PopSteer}, we conducted a controlled neuron manipulation study. Specifically, we manually deactivated the top-$K^\prime$ neurons identified by \algname{PopSteer} as most strongly associated with popularity (Cohen’s $d>1$) or unpopularity (Cohen’s $d<-1$).

Figure~\ref{fig:interpretation} illustrates how Gini Index evolves as these neurons are progressively turned off. When neurons contributing to popularity are deactivated (blue line), the Gini Index consistently decreases—highlighting their role in reinforcing popularity bias. Conversely, turning off neurons associated with unpopular items (orange line) leads to an increase in Gini Index, confirming their importance in enhancing exposure fairness. These findings demonstrate that \algname{PopSteer} not only mitigates popularity bias effectively but also provides actionable, interpretable insights into the neural basis of bias—enabling more transparent and controllable interventions.

\subsubsection{Ablation study}

To verify the effectiveness of controlled steering, we perform a comparative analysis with a baseline in which random Gaussian noise is added to neuron activations instead of targeted neuron steering. Specifically, we vary the number of neurons $\mathcal{N}$ subjected to perturbation, while fixing the hyperparameters $\alpha$ for \algname{PopSteer} and the standard deviation ($\xi$) for the Gaussian noise. Both hyperparameters ($\alpha$ and $\xi$) are tuned individually for each dataset to ensure that the resulting reduction in nDCG@10 does not exceed 10\% compared to the original unsteered model.

Table~\ref{tab:pop_vs_noise_core} presents the results of our ablation study, comparing \algname{PopSteer} to the noise-based baseline. We observe that \algname{PopSteer} consistently yields superior fairness outcomes relative to random Gaussian noise perturbation across both datasets. Specifically, while the noise-based method shows minimal variability and no meaningful increase in fairness metrics as we vary the number of perturbed neurons ($\mathcal{N}$), \algname{PopSteer} demonstrates a clear and predictable improvement in fairness with increasing $\mathcal{N}$. This indicates that the fairness improvements achieved by \algname{PopSteer} are directly attributable to targeted neuron selection rather than random perturbations.

\section{Conclusion}

This work introduces \algname{PopSteer}, a post-hoc method for interpreting and mitigating popularity bias in recommendation models using Sparse Autoencoder. By identifying and adjusting neuron activations linked to popularity signals, \algname{PopSteer} improves exposure fairness without sacrificing accuracy. Our experiments on public datasets demonstrate its effectiveness and reliability compared to existing baselines. Beyond performance, the method offers interpretability and fine-grained control, making it a practical and transparent solution for fairness-aware recommendation.

\bibliographystyle{ACM-Reference-Format}
\bibliography{ref}

\end{document}